\documentclass[twocolumn,aps,tightenlines,floatfix,showpacs]{revtex4}
\usepackage[dvips]{graphicx}
\usepackage{hhline}

\newcommand{\be}{\begin{equation}}
\newcommand{\ee}{\end{equation}}

\begin{document}
\title{A High-Performance Algorithm to Calculate Spin- and
Parity-Dependent Nuclear Level Densities}

\author{R.A. Sen'kov and M. Horoi}

\affiliation{Department of Physics, Central Michigan University, Mount Pleasant, Michigan 48859, USA}


\begin{abstract}
A new algorithm for calculating the spin- and parity-dependent shell model nuclear level 
densities using the moments method in the proton-neutron formalism is presented. A new, 
parallelized code based on this algorithm was developed and tested using up to 4000 cores 
% on one of NERSC supercomputers, 
for a set of nuclei from the $sd$-, $pf$-, and $pf + g_{9/2}$ - model spaces. 
By comparing the nuclear level densities at low excitation energy for a given
nucleus calculated in two model spaces, such as $pf$ and $pf + g_{9/2}$, one
can estimate the ground state energy in the larger model space, which
is not accessible to direct shell model calculations due to the unmanageable dimension. 
Examples for the ground state energies of for ${}^{64}$Ge and ${}^{68}$Se in 
the $pf + g_{9/2}$ model space are presented.
\end{abstract}

\maketitle

\section{Introduction}

Spin- and parity-dependent nuclear level densities represent an important ingredient 
for the theory of nuclear reactions. For example, the Hauser-Feschbach approach \cite{HF}
requires exact knowledge of nuclear level densities for certain quantum numbers $J^{\pi}$ of spin and
parity in the Gamow window of excitation energies around the
particle threshold \cite{rauscher97,moller09}. In most of the cases relevant
to nuclear astrophysics, where experimental information is not available, the reaction 
rates for medium and heavy nuclei can only be estimated using the Hauser-Feshbach approach.
Nuclear level densities are usually obtained using the back-shifted Fermi gas %(BSFG)
approximation \cite{bethe36,cameron,ericson}, which was improved over the years. More modern approaches to
the level densities based on the mean field were recently proposed by Goriely and 
collaborators \cite{gor-icomb-08,gor-fis-09}, and \cite{aberg09}. 
These approximations assume an independent particle model in a mean field that lacks 
information about the many-body correlations. These correlations can be included exactly if
one can fully diagonalize the many-body nuclear Hamiltonian, a task of increasing difficulty.
Alternatively, one can use Monte-Carlo techniques \cite{Eric97,Alhassid97,Langanke98,alhassid00,alhassid07,alhassid09}, 
or other methods of the statistical spectroscopy \cite{calvin-06-1,calvin-06-2}, including applications
to large shell-model spaces \cite{kota-96,huang-2000}.
%{\bf The applications to experimental data are given in many papers, see for example \cite{kota-96,huang-2000}.}

Most of these methods \cite{rauscher97,dorel-09,Eric97,Alhassid97,Langanke98,gor-ncomb-htm}
calculate the {\sl density of states} and later use a spin-weight factor
that includes an energy-dependent cut-off parameter
 to extract the spin-dependent {\sl nuclear level density}. Although there are recent
efforts to improve the accuracy of such parametrizations \cite{dorel-09}.
%{\bf Statistical spectroscopy supplies methods for calculating the energy-dependent spin cutoff parameter}. 
It was shown that the spin cut-off parameter has very large fluctuations at low excitation energy, when compared
with the shell model results \cite{nic8den}. Statistical spectroscopy provides
a path to a direct calculation of the spin cut-off parameter using a polynomial expansion for its
estimate \cite{wongbook}. This approach was recently investigated (see Fig. 2 of Ref. \cite{nic8den}), where it was shown
that although the smooth part of the energy dependence of the spin cut-off parameter can be
described reasonably well, significant fluctuations are still present in the low-energy regime. % (up to 15-20 MeV).
The quality of the results of this approach are mixed. Therefore, one would like to have a spin-projected
method of calculating nuclear level densities that is accurate and fast.
The parity is usually taken as equally distributed, although there are attempts \cite{alhassid00,mocelj-pm} 
to model the effect of the uneven parity-dependence of the level densities at the low excitation energies of
interest for nuclear astrophysics.

Recently, we developed a strategy \cite{jtpden,jdenrc,nic8den,dencom} of
calculating the spin- and parity-dependent shell model level
density. The main ingredients are: (i) extension of methods of
statistical spectroscopy \cite{french83,wongbook} by exactly
calculating the first and second moments for different
configurations at fixed spin and parity; (ii) exact decomposition of many-body
configurational space into classes corresponding to different
parities and number of harmonic oscillator excitations; (iii)
development of new effective interactions for model spaces of
interest starting with the $G$-matrix \cite{gmat} and fixing/fitting
monopole terms and/or linear combinations of two-body matrix
elements to experimental data; and (iv) an accurate estimate of the
shell model ground state (g.s.) energy.  The calculation of the latest ingredient is generally 
as time consuming as the previous three. One can minimize this effort using the 
exponential convergence method suggested and applied in 
Refs. \cite{ecm,aecm1,aecm2}, and/or the recently developed projected configuration
interaction method \cite{pci1,pci2}. In reverse, one can envision using some 
knowledge about the level density to extract the g.s. energy. 
This idea is not new (see, for example, Refs. \cite{wongbook,chang-71,kar-97,choubey-98}).
However, we propose new algorithm that extracts the g.s. energy for a large model space by
comparing the level density with that obtained in a reduced model space that can be
exactly solved.

%These approximations assume an independent particle model in a mean-field, which lack 
%information about the many body correlations. These correlations can be included exactly if
%one can fully diagonalize the many body nuclear Hamiltonian, a task of increased difficulty.
%Alternatively, one can use Monte-Carlo techniques \cite{12}, or methods of the statistical
%spectroscopy\cite{6}. We recently developed a methodology \cite{dencom,2,3,4} of calculating 
%the spin and parity dependent shell model nuclear level densities, which
%is a very useful input to the Huaser-Feshbach theory
%for calculating reaction rates for nuclear astrophysics \cite{5}.
The techniques described in this article are based on nuclear statistical spectroscopy \cite{wongbook}. 
We calculate the configuration spin and parity projected moments of the
nuclear shell model Hamiltonian, which can be further used to 
obtain an accurate description of the nuclear level density up to about 15 MeV
excitation energy. Therefore, our methodology does not require any  spin-cut-off parameter.
One should mention that some of the more recent Monte Carlo approaches for level densities can also 
use direct spin projection techniques \cite{alhassid07}.

The article is organized as follows.
In Sec. II the fixed spin- and parity-dependent configuration moments method is revisited. 
The method allows one to trace such quantum numbers as parity 
and angular momentum explicitly. The extension of the algorithm to the proton-neutron formalism is 
discussed in Sec. III. 
Section IV is devoted to the results of the moments method, which are compared to exact shell model results and the results 
from Hartree-Fock-Bogoliubov plus combinatorial method.  In Sec. IV we also present our new algorithm
to extract the g.s. energy by comparing level density in related model spaces. 
Section V is devoted to conclusions and future prospects of the moments method.

\section{Spin- and Parity-dependent configuration moments method}

In this work we closely follow the approach proposed 
in Refs. \cite{jtpden,jdenrc}. According to this approach one can calculate 
the nuclear level density $\rho$ %with a given set of quantum numbers 
as a function of excitation energy $E$ in the following way:
\be \label{rho1}
\rho(E,\alpha) = \sum_\kappa D_{\alpha \kappa } \cdot G_{\alpha \kappa} (E).
\ee
Here, $\alpha = \{n, J, T_z, \pi \}$ is a set of quantum numbers, 
where $n$ is the number of particles (protons and neutrons), $J$ is total spin, 
$T_z$ is isospin projection, and $\pi$ is parity. 
$\kappa$ represents a configuration of $n$ particles
distributed over $q$ spherical single-particle orbitals. Each 
configuration $\kappa$ is fixed by a set of occupation 
numbers $ \kappa=\{\kappa_1, \kappa_2,... \; , \kappa_q \}, $ 
where $\kappa_j$ is the number of particles occupying the spherical single-particle 
level $j$. 
Each configuration has a certain number of particles, isospin projection,
and parity. The sum in Eq. (\ref{rho1}) spans all possible 
configurations corresponding to the given values of $n,T_z$, and $\pi$.
The dimension $D_{\alpha \kappa}$ equals the number of many-body states with given 
$J$ that can be built for a given configuration $\kappa$. The function $G_{\alpha \kappa}$ 
is a finite-range Gaussian defined as in Ref.\cite{jtpden}:
\begin{eqnarray}\label{frg}
G_{\alpha \kappa}(E)=G(E+E_{gs}-E_{\alpha \kappa},\sigma_{\alpha \kappa}),\\
G(x,\sigma)= N \cdot \left\{ 
\begin{array}{ll}
\mbox{exp}\left( -x^2/2\sigma^2 \right) &, \; \; |x| \leq \eta \cdot \sigma \\
0 &, \; \; |x|> \eta \cdot \sigma \\
\end{array} \right. ,
\end{eqnarray}
where $E_{\alpha \kappa}$ and $\sigma_{\alpha \kappa}$ are the fixed-$J$ centroids 
and widths, which will be defined later, $E_{gs}$ is the ground state energy, $\eta$ 
is the cut-off parameter, and $N$ is the normalization factor corresponding to the 
following condition: $\int_{-\infty}^{+\infty} G(x,\sigma) dx = 1$.
In this work we treat $\eta$ as a free parameter. From previous works
(see for example \cite{nic8den}) we know that the cut-off parameter is $\eta \sim 3$.
We can slightly variate the value of $\eta$ to achieve a better 
description of the nuclear level density, see Fig. (\ref{eta}).

\begin{figure*}
\includegraphics[width=0.45\textwidth]{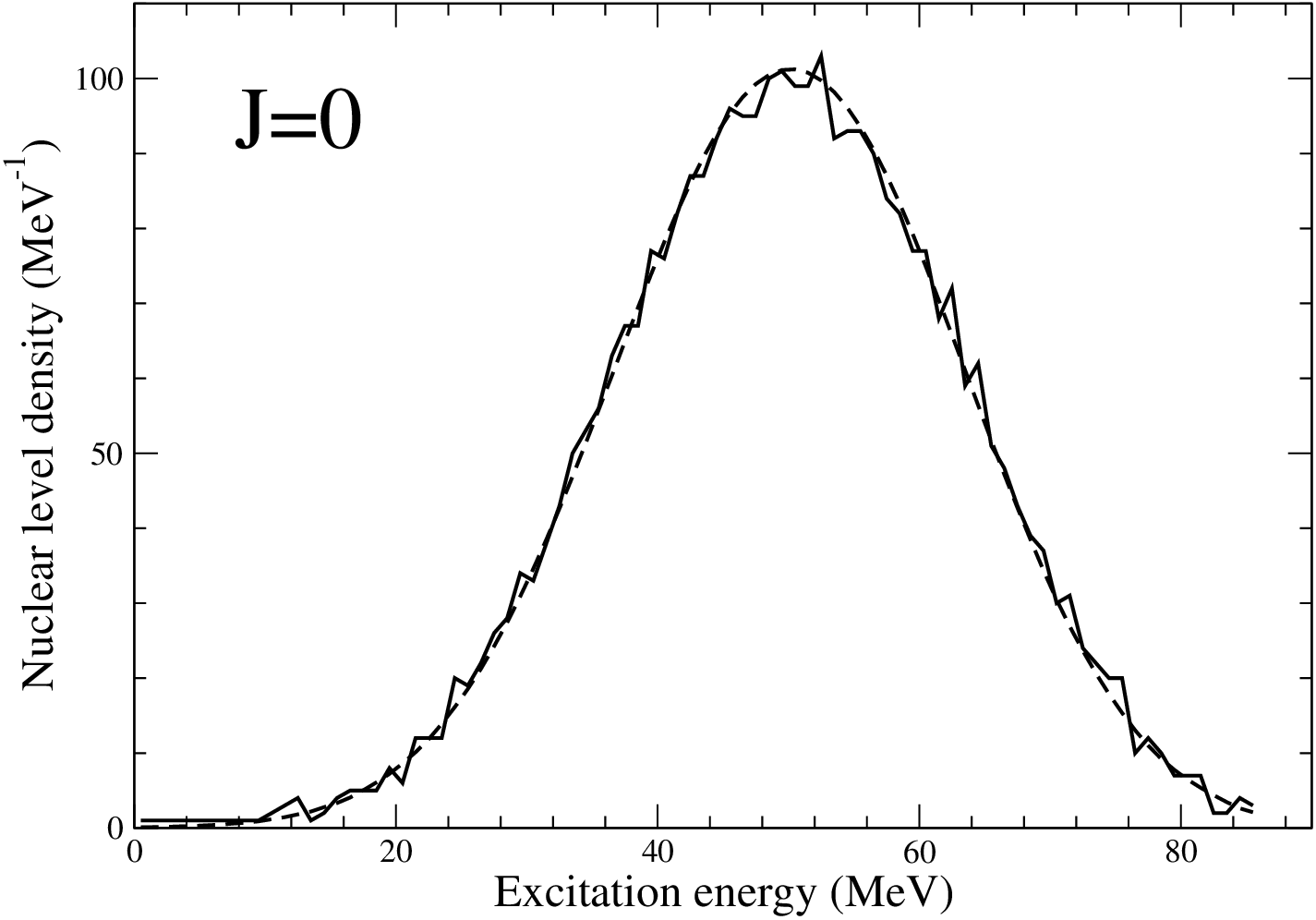}
\hfill
\includegraphics[width=0.45\textwidth]{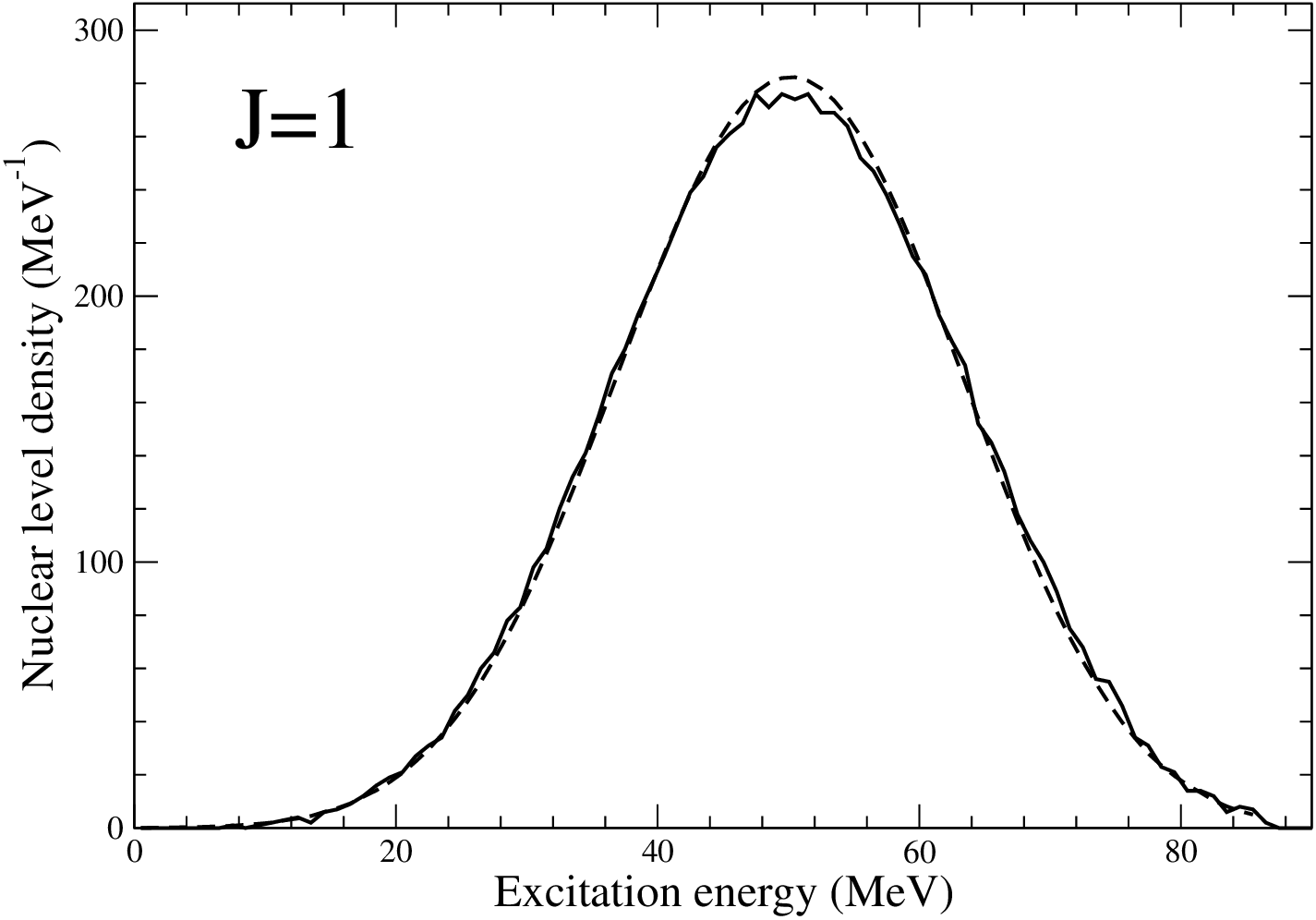}
\\

\vspace{1.0cm}

\includegraphics[width=0.45\textwidth]{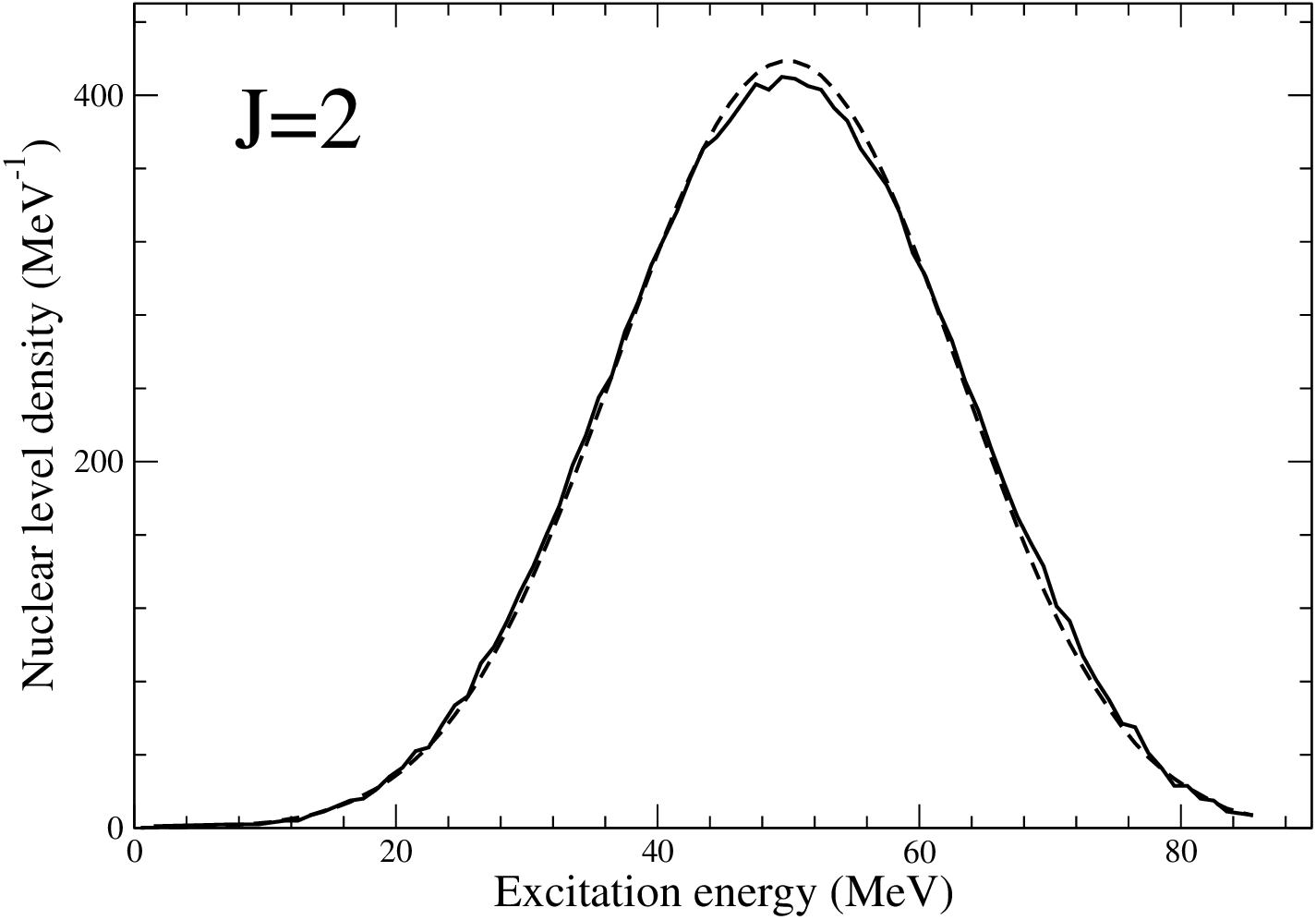}
\hfill
\includegraphics[width=0.45\textwidth]{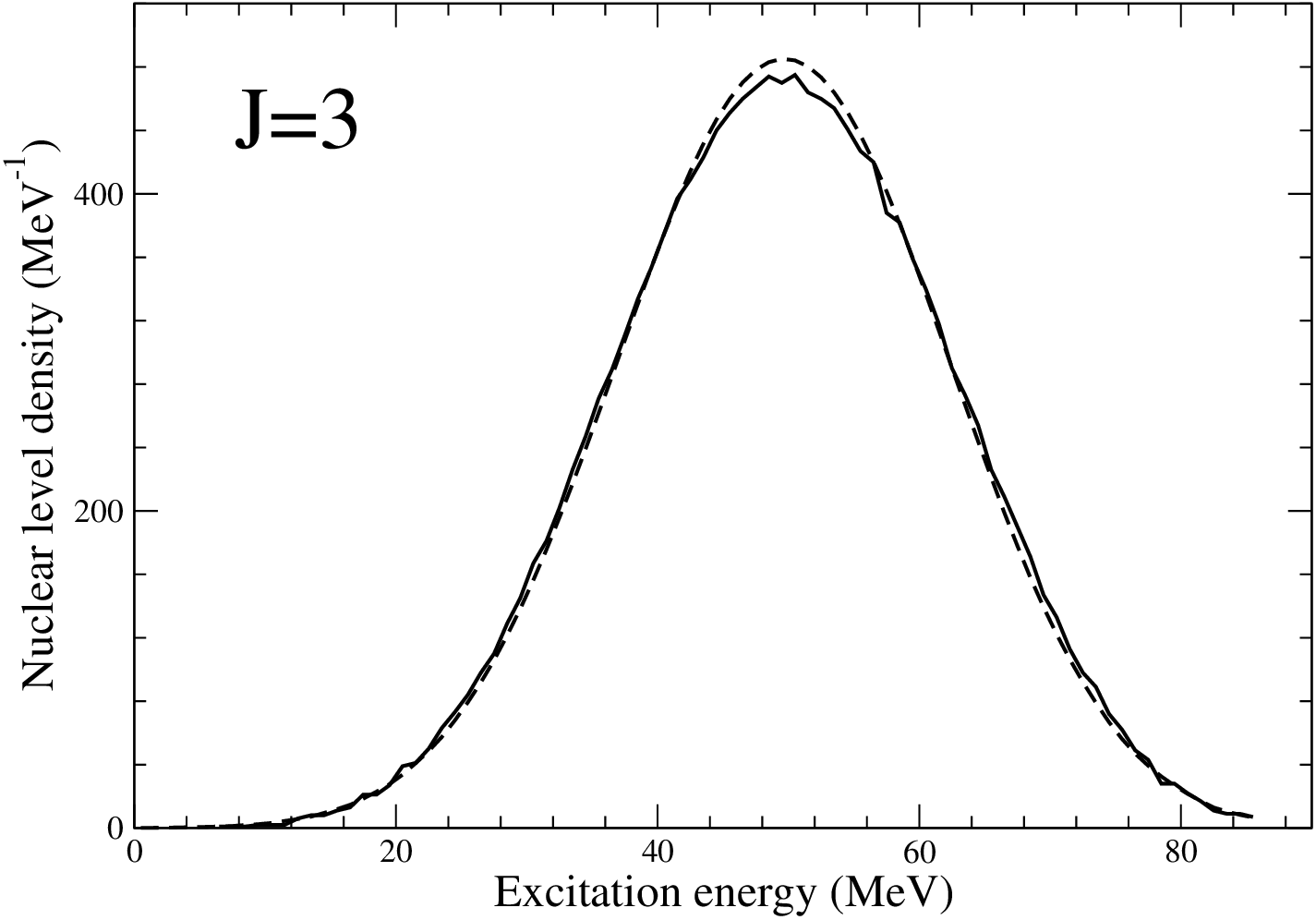}
\\

\caption{${}^{28}$Si, parity=+1. Comparison of nuclear level densities 
between exact shell model (solid line) and moments method (dashed line). 
Cut-off parameter $\eta=2.8$, interaction: USD, $sd$-shell. }\label{si28}
\end{figure*}

The $J$-dependent moments method provides a good description of the exact 
$J$-dependent shell-model level density. 
Figure \ref{si28} presents the results for ${}^{28}$Si in the  
$sd$-shell for different values of spin $J$ and positive parity. 
Figures \ref{fe52} and \ref{cr52} present results for ${}^{52}$Fe and ${}^{52}$Cr 
nuclei in the $pf$-shell. % for $J^{\pi}=0^+$. 
Similar results were obtained for the density of states using the general moments method 
(see examples in Refs. \cite{nic8den,wongbook,brody81}). 
 A very important ingredient for our method is an accurate knowledge of 
the ground state energy $E_{gs}$.
It is also important to investigate the sensitivity of the results to the the cut-off parameter $\eta$, 
and find optimal values for it.
These issues will be discussed in more detail in Sec. IV.

Let us define now the fixed-$J$ centroids and widths from Eq. (\ref{frg}).
To calculate them for a two-body Hamiltonian,
\be
\label{h}
H = \sum_i \epsilon_i a^\dag_i a_i + \frac{1}{4} \sum_{i j k l} V_{i j k l}
a^\dag_i a^\dag_j a_l a_k,
\ee
one has to calculate traces of the first and second power of this Hamiltonian,
Tr$[H]$ and Tr$[H^2]$, for each configuration $\kappa$:
\begin{eqnarray}
E_{\alpha \kappa} = \left< H \right>_{\alpha \kappa}, \\
\sigma_{\alpha \kappa} = \sqrt{ \left< H^2 \right>_{\alpha \kappa} 
- \left< H \right>^2_{\alpha \kappa} },
\end{eqnarray}
where
\begin{eqnarray}
\label{tr1}
\left< H \right>_{\alpha \kappa} = 
\mbox{Tr}^{(\alpha \kappa)}[H]/D_{\alpha \kappa},\\
\label{tr2}
\left< H^2 \right>_{\alpha \kappa} = 
\mbox{Tr}^{(\alpha \kappa)}[H^2]/D_{\alpha \kappa}.
\end{eqnarray}
Here the symbol of trace $\mbox{Tr}^{(\alpha \kappa)}[\cdots]$ means the sum of all diagonal
matrix elements, as $\sum \left< \nu, J | \cdots | \nu, J \right>$, over all many-body 
states $\left|\nu, J \right>$
belonging to the given configuration $\kappa$ and having a certain set of quantum
numbers $\alpha$, including spin $J$.
Technically, it is more convenient to derive these traces in a basis with a fixed 
spin projection $\left| \nu, M_z \right>$, $\mbox{Tr}^{(M_z)}[\cdots]$, rather than in the basis with 
fixed total spin $\left| \nu, J \right>$, $\mbox{Tr}^{(J)}[\cdots]$. $J$-traces can be easily expressed 
through the $M_z$-traces, given the spherical symmetry of the Hamiltonian:
\be
\label{mztoJ}
\mbox{Tr}^{(J)} [\cdots] 
= \mbox{Tr}^{(M_z)} [\cdots]{\Big \vert}_{{}_{M_z=J}} -
\mbox{Tr}^{(M_z)} [\cdots] {\Big \vert}_{{}_{M_z=J+1}}.
\ee
For simplicity, in Eq. (\ref{mztoJ}) we omitted all quantum numbers, except the projection $M_z$ and 
the total spin $J$. 

\begin{figure}
\includegraphics[width=0.48\textwidth]{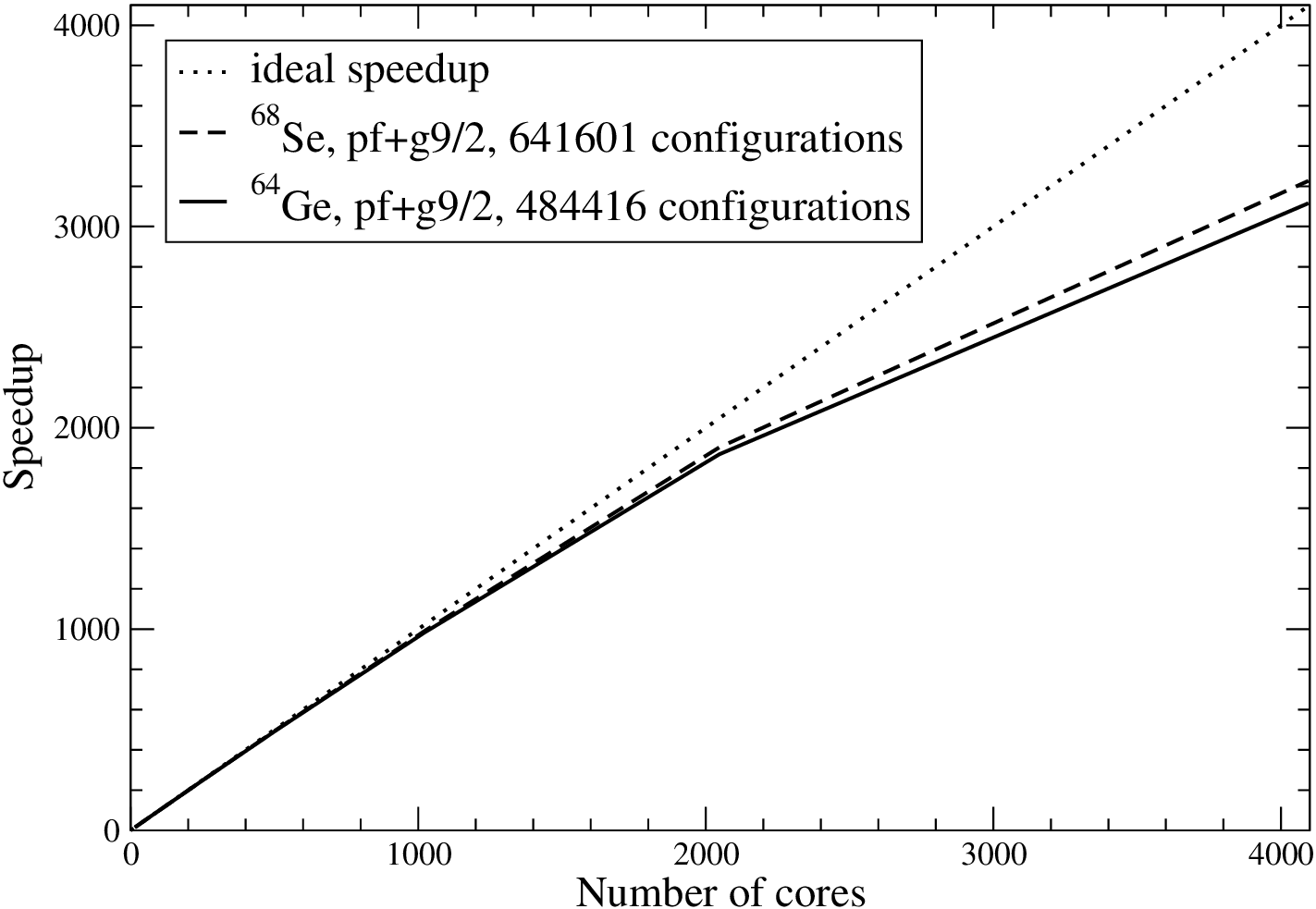}
\caption{Speedup is defined as $T_1/T_n$, where $T_n$ is the calculation time,
when $n$ processors were used. These calculations were performed on FRANKLIN supercomputer at the National 
Energy Research Scientific Computing Center (NERSC) \cite{nersc}.}\label{spup}
\end{figure}

Hereafter we use the label $\alpha$ to denote a set of quantum numbers that includes
either the
fixed $M_z$ or the fixed $J$, keeping in mind that Eq. (\ref{mztoJ}) 
can always connect them. 
In every important case we will point out which set of quantum numbers was used. 
Following the approach of Ref. \cite{jaq79} (a similar method can be found
in Ref. \cite{verb-81}), we obtain the following
expressions for the traces from Eqs. (\ref{tr1}) and (\ref{tr2}):
\begin{eqnarray}
\label{tr1c}
\mbox{Tr}^{(\alpha \kappa)}[H] = \sum_i \epsilon_i  D^{[i]}_{\alpha \kappa}
+ \sum_{i<j} V_{ijij}D^{[ij]}_{\alpha \kappa},\;\;\;\;\\
\nonumber
\mbox{Tr}^{(\alpha \kappa)}[H^2] = \sum_i \epsilon^2_i  D^{[i]}_{\alpha \kappa} + \\
\nonumber
+ \sum_{i<j} \left[ 2 \epsilon_i \epsilon_j +  2 (\epsilon_i+\epsilon_j) V_{ijij} + \sum_{q<l}
V^2_{ijql} \right] D^{[ij]}_{\alpha \kappa} + \\
\nonumber
+\sum_{(i<l)\ne l} \left[ \sum_q \left( 2V_{liiq}V_{ljjq}-V^2_{ijql}\right) 
+ 2 \epsilon_l V_{ijij}\right] D^{[ijl]}_{\alpha \kappa} + \\
\label{tr2c}
+ \sum_{(i<j)\ne(q<l)} \left[ V^2_{ijql}+V_{ijij}V_{qlql}-4V_{qiil}V_{qjjl}\right]
D^{[ijql]}_{\alpha \kappa},\;\;\;\;
\end{eqnarray}
where $i,j,l,$ and $q$ are single-particle states with certain 
spin projections and possible occupation numbers equal to 0 or 1.
Notice that the single-particle orbitals we used to define 
the configurations in Eq. (\ref{rho1}), can host all particles with all possible 
spin projections corresponding to the orbital's spin.
$D^{[i]}_{\alpha \kappa} = \mbox{Tr}^{(\alpha \kappa)}[a^\dag_i a_i]$ 
can be interpreted as a number of many-body states with fixed projection 
$M_z$ (if we consider $M_z$-traces) and the single-particle 
state $i$ occupied, which can be constructed for the configuration $\kappa$, 
$D^{[ij]}_{\alpha \kappa}= \mbox{Tr}^{(\alpha \kappa)}[a^\dag_i a^\dag_j a_j a_i]$,
$D^{[ijq]}_{\alpha \kappa} = \mbox{Tr}^{(\alpha \kappa)}[a^\dag_i a^\dag_j a^\dag_q a_q a_j a_i]$, and so on.
These $D$-structures were called propagation functions in Ref. \cite{jaq79}.
For completeness, we repeat here the recipe used to calculate them.
One can show \cite{jaq79} that
\be
\label{rec}
D^{[r_1 r_2 \cdots r_s]}_{\alpha \kappa} = \sum_{s \leq t\leq n}(-1)^{t-s}\sum_{t_1+\cdots +t_s=t} 
D_{\alpha' \kappa'},
\ee
where all $t_i$ are integers, 
configuration $\kappa'=\{\kappa'_1, \kappa'_2,... \; , \kappa'_q \}$ 
can be derived from the original configuration $\kappa=\{\kappa_1, \kappa_2,... \; , \kappa_q \}$
by removing $t$ particles corresponding to the single-particle states $r_1, r_2, \cdots r_s$.
Formal expression for the new $\kappa'$ configuration can be written as follows:
\be
\kappa'_j=\kappa_j - \sum_{i \; (r_i \in j)} t_i,
\ee
where the sum includes only 
those values of $i$ for which the corresponding single-particle state $r_i$ belongs
to the single-particle level $j$. 
We also assume that all the occupation numbers $\kappa'_j$ must be positive, which imposes certain
restrictions on the possible values of the amplitudes $t_i$. For every new configuration $\kappa'$ one 
can easily define new quantum numbers, $\alpha'=\{n' M'_z T'_z \pi' \}$, entering Eq. (\ref{rec}).
Examples are the new number of particles $n'=n-t$ and the new spin projection:
\be
M'_z=M_z-t_1 m_{r_1}-t_2 m_{r_2}-\cdots-t_s m_{r_s},
\ee
where $m_{r_i}$ is the $M_z$ projection of the single-particle state $r_i$.
The new isospin $T'_z$ and parity $\pi'$ are defined similarly. 

\section{The moments method algorithm in the proton-neutron formalism}

Let us describe some technical features of the algorithm we developed for the nuclear
level density calculation. First of all, we treat protons and neutrons separately, that is, 
the basis of many-body wave functions are represented by a product of proton and neutron 
parts:
\be
\label{bwf}
|\nu, M_z\rangle = |\nu_p, M^{(p)}_{z}\rangle \cdot |\nu_n, M^{(n)}_{z}\rangle,
\ee
where $M^{(p)}_{z}+M^{(n)}_{z}=M_z$. Thus, the wave functions (\ref{bwf}) have
fixed isospin projection $T_z$, but do not have a fixed isospin $T$. As 
we already mentioned, it is more convenient to use the basis of wave functions with fixed 
spin projection $M_z$, rather than one with fixed spin $J$.

One can gain essential advantages from such a proton-neutron separation of the basis. 
One of them is connected to the number of configurations that appear
in the sum of Eq. (\ref{rho1}). Naturally, the number of configurations with fixed
$T_z$ is much greater than the number of configurations with fixed isospin. The large
number of configurations allows the use of many-cores computers with greater efficiency. 
In other words, the calculation of the sum in Eq. (\ref{rho1}) with a larger number of
configurations
can be more efficiently distributed on a larger number of processors. Figure \ref{spup}
presents the speedup (calculation speed gain) as a function of the number of used processors. 
One can see that the case with the larger number of configurations, ${}^{68}$Se, scales better 
than the case with the lower number of configurations, ${}^{64}$Ge. Up to 2000 cores the 
speedup is almost perfect (the dotted line presents an ideal speedup). At this point the
calculation time is about of 1-2 minutes and further improvement is
hardly achieved. 

Another significant advantage of the proton-neutron formalism is the new algorithm of calculation of the dimensions 
$D_{\alpha \kappa}$, $D^{[i]}_{\alpha \kappa}$, $D^{[i j]}_{\alpha \kappa}$, and so on.
Because of the proton-neutron separation one can calculate all proton and neutron dimensions separately. 
Later, the dimensions we are interested in can be easily constructed 
from the proton and neutron parts using the following convolution
\be
\label{drel}
D_{M_z \kappa} = \sum_{M^{(p)}_z+M^{(n)}_z=M_z} D_{M^{(p)}_z \kappa_p} \cdot D_{M^{(n)}_z \kappa_n},
\ee
where, instead of the whole set of quantum numbers $\alpha$, only the spin projection $M_z$ was printed out.
$\kappa_p$ and $\kappa_n$ are the proton and neutron parts of the configuration $\kappa$.
Equation (\ref{drel}) can be easily applied to all types of dimensions, $D^{[\cdots]}_{\alpha \ldots}$,  
we have shown in the formalism of Sec. II. The advantage comes
from the fact that one can calculate and keep in memory all proton and neutron dimensions,  
$D_{M^{(p)}_z \kappa_p}$ and $D_{M^{(n)}_z \kappa_n}$, for all possible projections $M^{(p)}_z$ and $M^{(n)}_z$,
and for all possible configurations $\kappa_p$ and $\kappa_n$.
Afterwards, using Eqs. (\ref{drel}) and (\ref{rec}), one can very quickly calculate all the dimensions: 
$D_{\alpha \kappa}$, $D^{[i]}_{\alpha \kappa}$, $D^{[i j]}_{\alpha \kappa}$, and so on, for all $M_z$ and $J$.

One more technical detail, which allows a significant speed up of the algorithm, is  that using the proton neutron separation one can avoid  multiple computations of the most time consuming structures, such as $D^{[i j q l]}_{\alpha \kappa}$. Let us consider a case when all four single-particle states $\{i j q l \}$ are protons. One can then use an equation similar to Eq. (\ref{drel}):
\be
\label{ndrel}
D^{[i j q l]}_{M_z \kappa} = \sum_{M^{(p)}_z+M^{(n)}_z=M_z} D^{[i j q l]}_{M^{(p)}_z \kappa_p} \cdot D_{M^{(n)}_z \kappa_n}.
\ee
For all configurations $\kappa$ that have the same proton parts $\kappa_p$ one will have to recalculate $D^{[i j q l]}_{M^{(p)}_z \kappa_p}$  for each neutron configuration. Alternatively, one can calculate $D^{[i j q l]}_{M^{(p)}_z \kappa_p}$ only once, and store them in memory. That strategy, however, will require a large amount of storage. More efficiently, one can only store the contributions of the  $D^{[i j q l]}_{\alpha \kappa}$ structures to the width,  Eq. (\ref{tr2c}), that is, one can only store  the following structures,
$$
T_{M^{(p)}_z \kappa_p}=\sum_{(i<j)\ne(q<l)} \left[ V^2_{ijql}+V_{ijij}V_{qlql} \right.
$$
\be \label{prc6}
\left. - 4V_{qiil}V_{qjjl}\right] D^{[ijql]}_{M^{(p)}_z \kappa_p},
\ee
where all single-particle states are protons. Thus, instead of using Eq. (\ref{ndrel}) one can calculate the contribution to the width directly via the convolution, 
\be
\mbox{Tr}^{(\alpha \kappa)}[H^2] = \dots + \sum_{M^{(p)}_z+M^{(n)}_z=M_z} T_{M^{(p)}_z \kappa_p} \cdot D_{M^{(n)}_z \kappa_n},
\ee
which is very similar to Eqs. (\ref{drel}) and (\ref{ndrel}).
As one can see the new approach avoids multiple calculations of $D^{[i j q l]}_{M^{(p)}_z \kappa_p}$. Storing the structures  
Eq. (\ref{prc6}) may significantly speed up the algorithm for large cases, such as ${}^{68}$Se in $pf+g_{9/2}$ model space. The downside is that the calculation of the structures $T_{M^{(p)}_z \kappa_p}, T_{M^{(n)}_z \kappa_n}$ does not always scale well on a large number of cores, since the number of these structures is much smaller than the total number of configurations.

\begin{table}[ht]
\begin{center}\label{ttime}
\begin{tabular}{||c|c|c|c||}
\hhline{|t:=:=:=:=:t|}
\hhline{||---||}
\rule{0cm}{0.33cm} Element & Space & Total dim & Elapsed time (sec)\\
\hhline{||----||}
\rule{0cm}{0.33cm} ${}^{70}$Br & $pf+g_{9/2}$ & $10^{15}$ & $1.07 \cdot 10^4$\\
\hhline{||----||}
\rule{0cm}{0.33cm} ${}^{68}$Se & $pf+g_{9/2}$ & $10^{15}$  & $1.03 \cdot 10^4$\\
\hhline{||----||}
\rule{0cm}{0.33cm} ${}^{64}$Ge & $pf+g_{9/2}$ & $10^{14}$ & $0.76 \cdot 10^4$\\
\hhline{||----||}
\rule{0cm}{0.33cm} ${}^{60}$Zn & $pf$ & $10^{11}$ & 37.4\\
\hhline{||----||}
\rule{0cm}{0.33cm} ${}^{52}$Fe & $pf$ & $10^{10}$ & 13.6\\
\hhline{||----||}
\rule{0cm}{0.33cm} ${}^{28}$Si & $sd$ & $10^{6}$ &0.7\\
\hhline{|b:=:=:=:=:b|}
\end{tabular}
\caption{Elapsed times of nuclear level density calculations 
(for all $J$, positive parity) with the moments method code. The calculations 
were done on a 16 cores machine with 2.8 GHz CPU frequency. \\
\bigskip }\label{tb1}
\end{center}
\end{table}

Table \ref{tb1} presents calculation times for different nuclei calculated in different shell-model spaces.
The calculations were done on a 16 cores machine with 2.8 GHz CPU frequency. One core ("master") 
distributed all the work between the other 15 cores ("slaves"). One can emphasize here that the listed times
correspond to calculations of the nuclear level densities for all $J$ and for positive parity.
For the case of ${}^{68}$Se the largest $m$-scheme dimension is about $10^{15}$.  For each $J$ the $m$-scheme
dimensions vary from $10^{12}$ to $10^{14}$, which makes direct 
diagonalization impossible. Using the moments method and our algorithm we are able to calculate 
the shapes of nuclear densities for ${}^{68}$Se in less than three hours on a 16 cores machine. 
If the number of processors reaches 1000 then one needs only a few minutes to complete the calculation.

\begin{figure}
\includegraphics[width=0.48\textwidth]{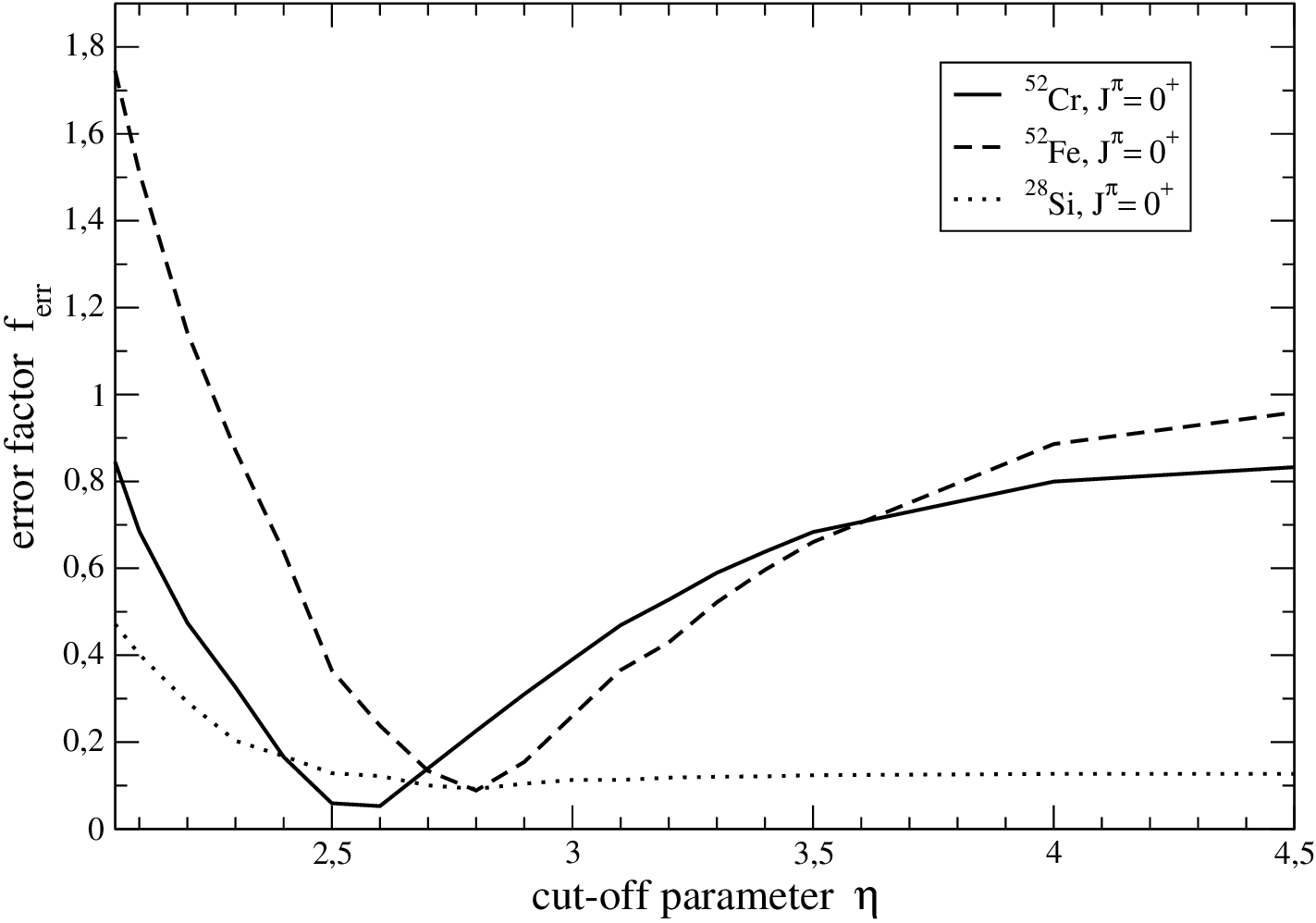}
\\
%\caption{${}^{52}$Fe and ${}^{52}$Cr }\label{eta}
\caption{Error factor $f_{err}$ as function of the cut-off parameter $\eta$. }\label{eta}
\end{figure}

\begin{figure*}
\includegraphics[width=0.47\textwidth]{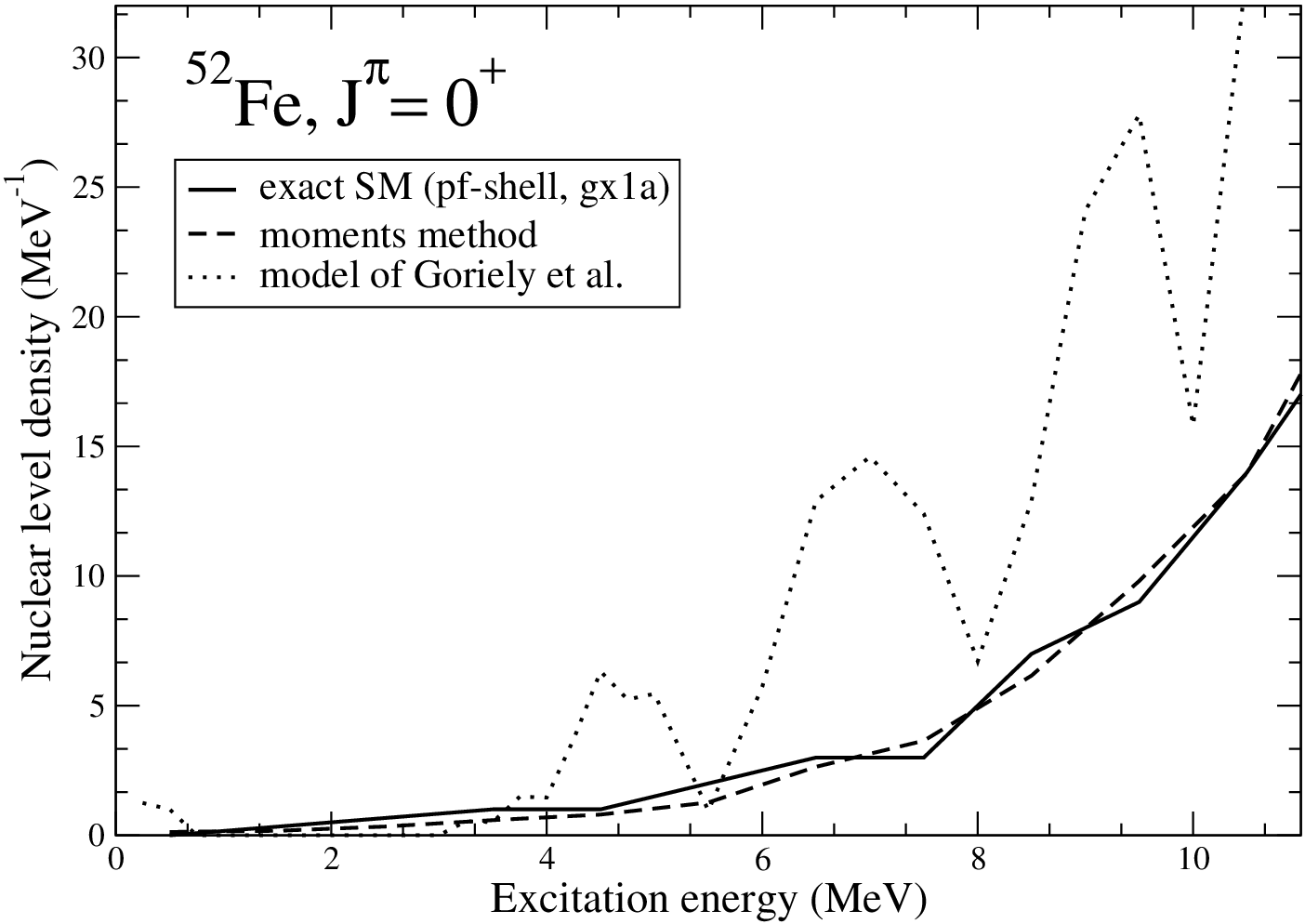}
\hfill
\includegraphics[width=0.47\textwidth]{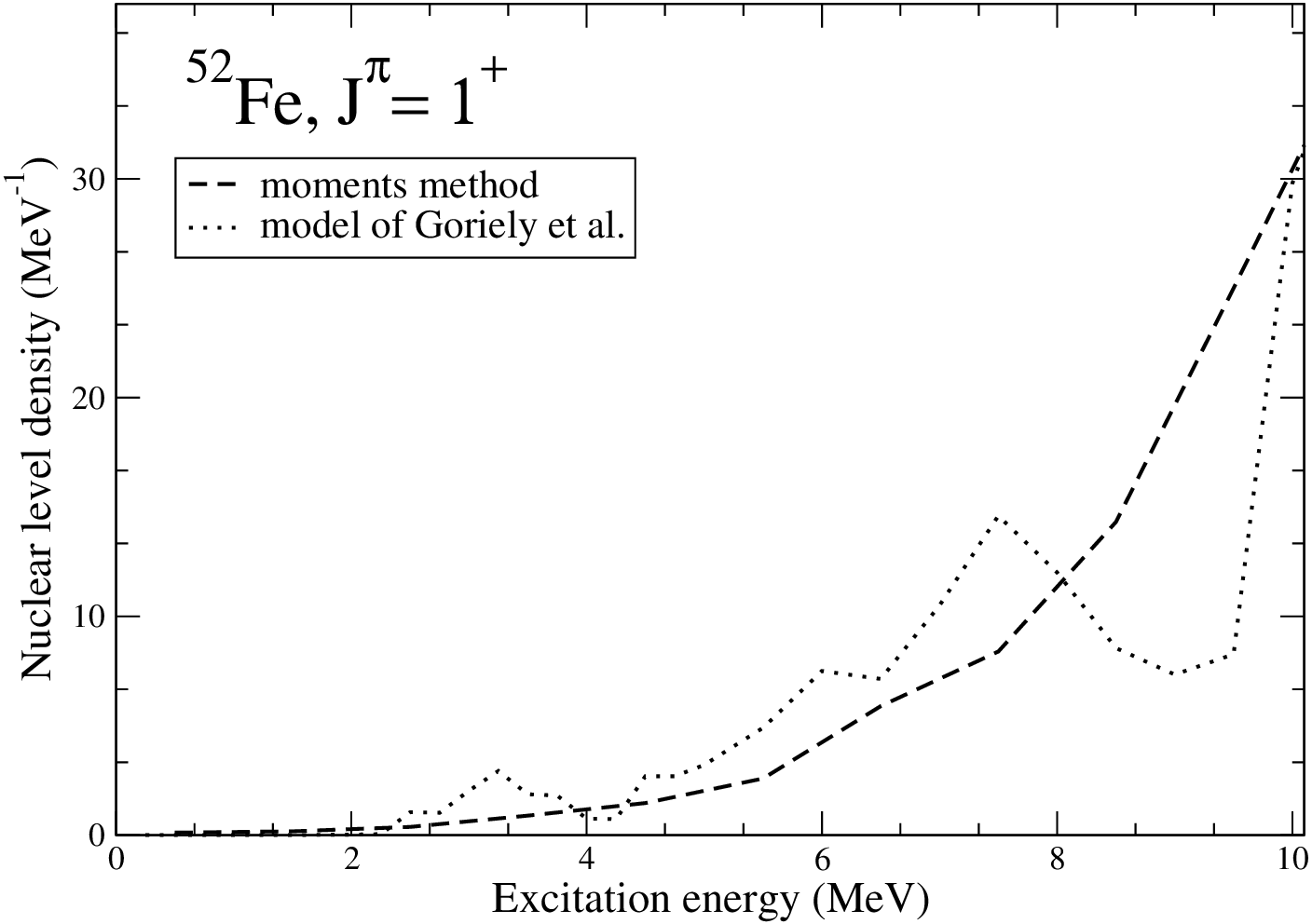}
\\

\caption{${}^{52}$Fe, parity=+1. Comparison of nuclear level densities 
between exact shell model (solid line) and moments method (dashed line). 
Cut-off parameter $\eta=2.6$, interaction: GXPF1A, $pf$-shell. }\label{fe52}

\end{figure*}

\begin{figure*}[h]

\includegraphics[width=0.47\textwidth]{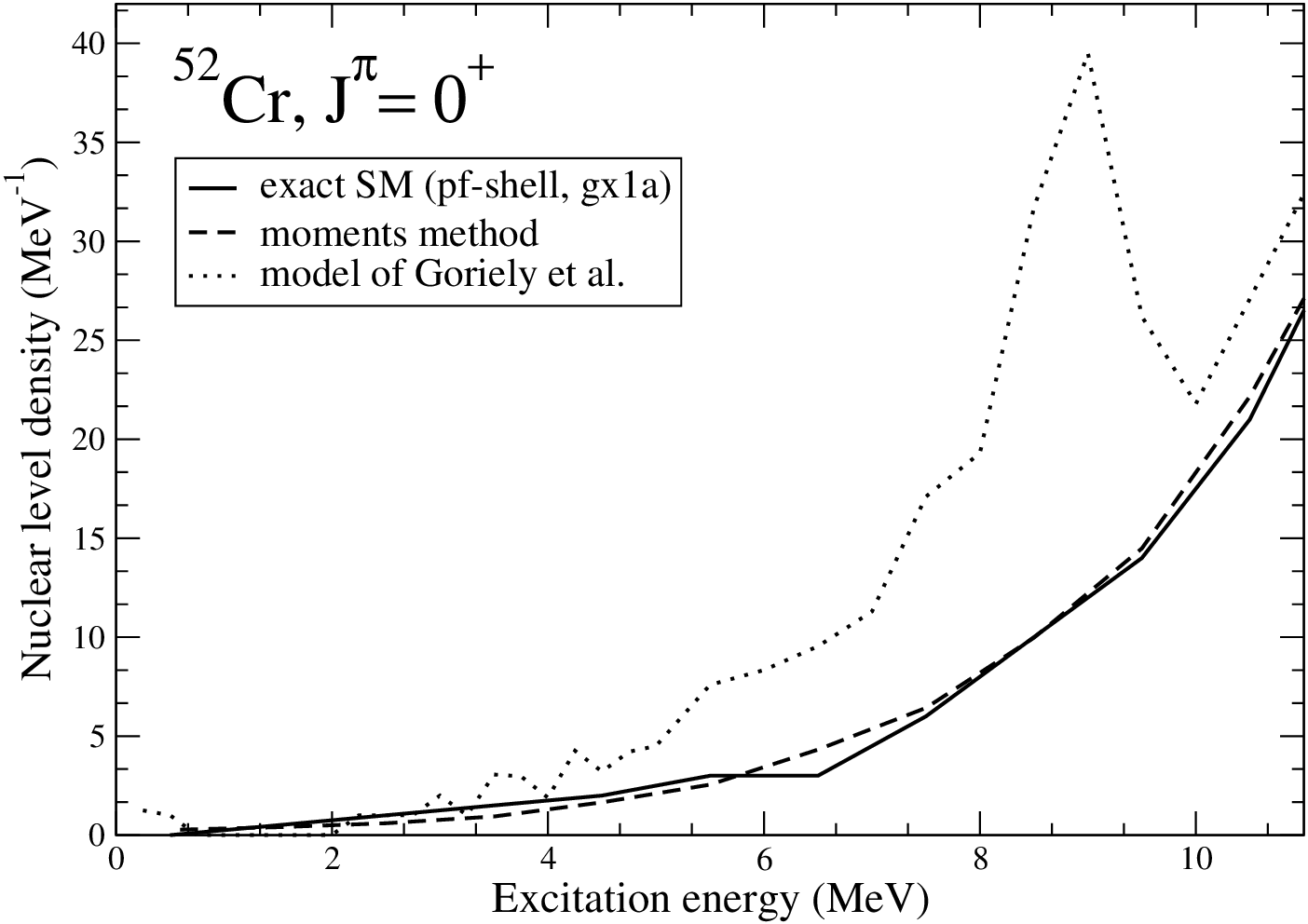}
\hfill
\includegraphics[width=0.47\textwidth]{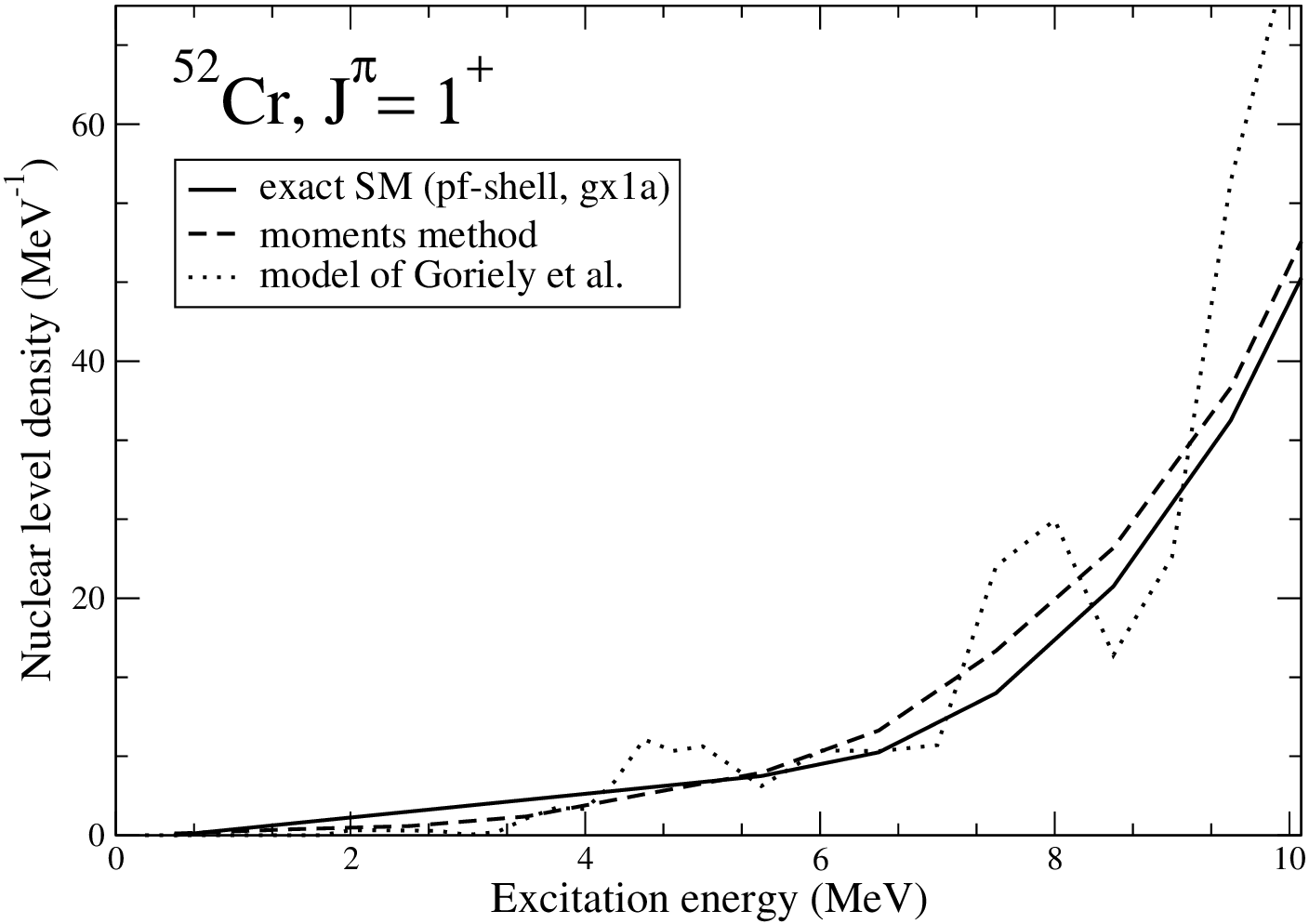}
\\

\vspace{1.3cm}

\includegraphics[width=0.47\textwidth]{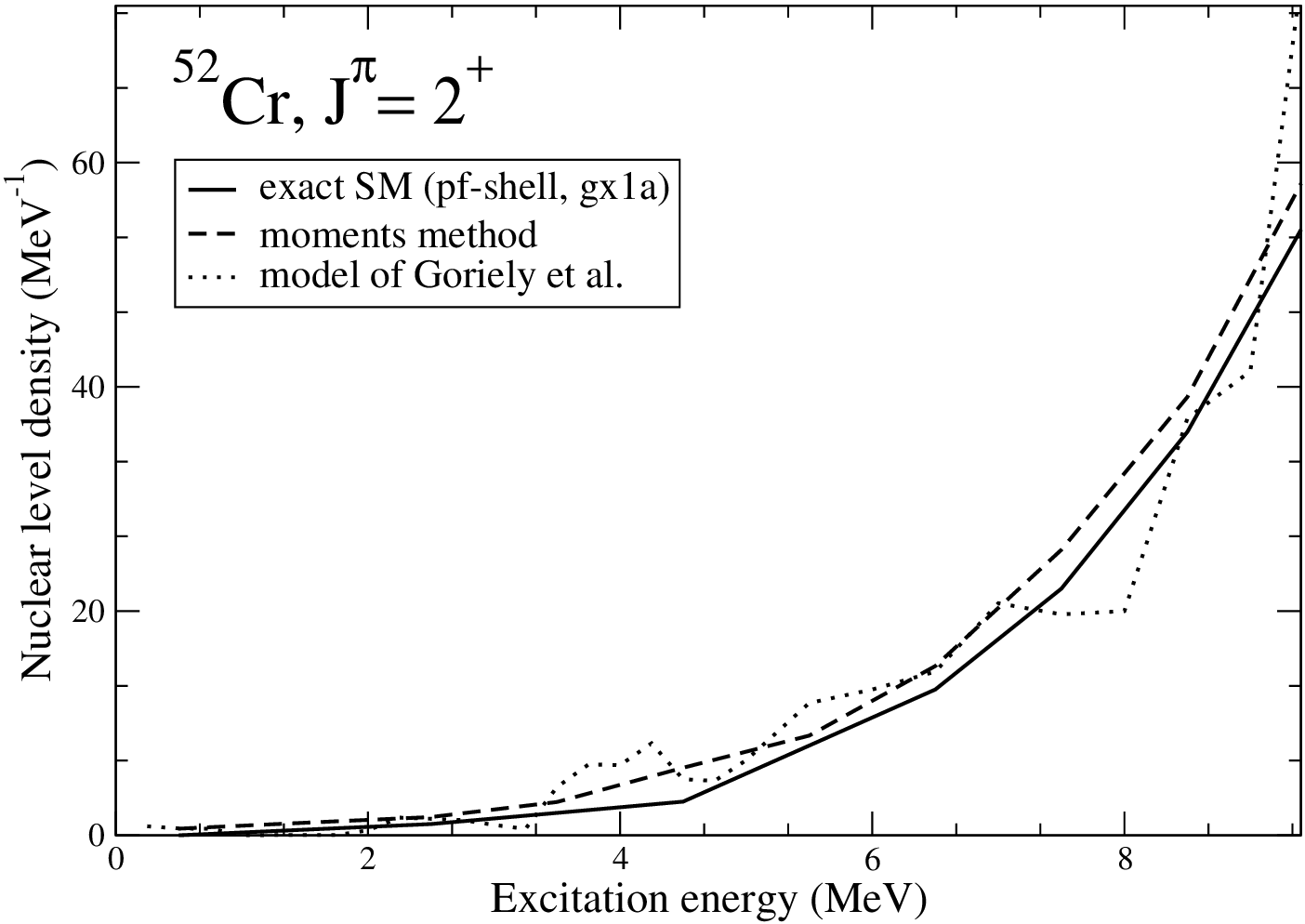}
\hfill
\includegraphics[width=0.47\textwidth]{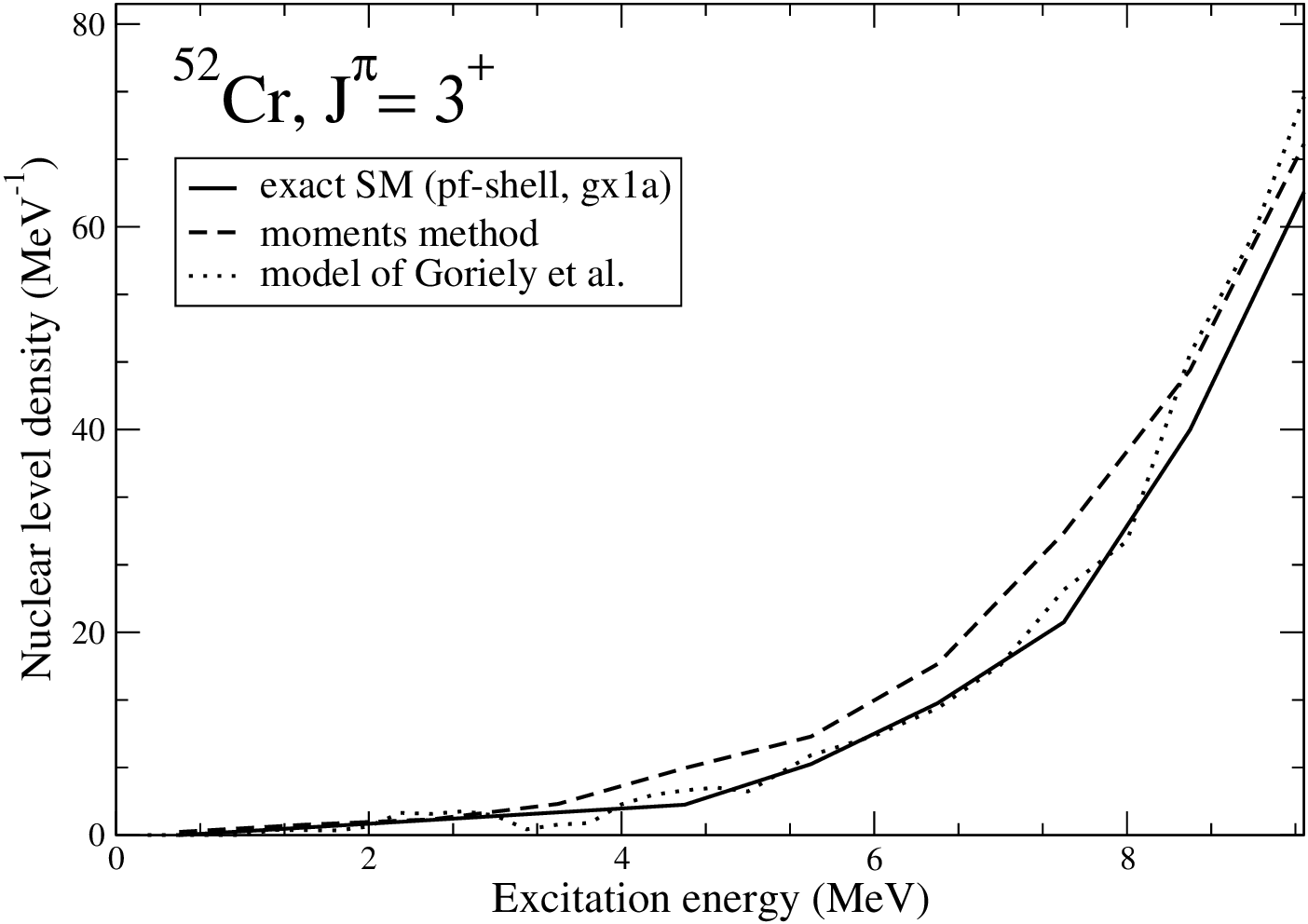}
\\

\caption{${}^{52}$Cr, parity=+1. Comparison of nuclear level densities 
between exact shell model (solid line) and moments method (dashed line). 
Cut-off parameter $\eta=2.6$, interaction: GXPF1A, $pf$-shell.\\
\bigskip }\label{cr52}

\end{figure*}

\section{Results}

%Part I. Test and comparison with Goriely's data.\\
%Let us first demonstrate that the moments method works. 
As a first example
we consider the nuclear level densities of ${}^{28}$Si in the $sd$-shell model 
space, for which we use the USD interaction \cite{usd}. Figure \ref{si28} presents
the comparison of the exact nuclear level densities of different spins (solid lines) with
those obtained with the moments methods (dashed lines). Equations (\ref{rho1}) and (\ref{frg}) require
the knowledge of the ground state energy $E_{gs}$ and the cut-off parameter $\eta$.
While the ground state energy of ${}^{28}$Si can be calculated in this case using 
the standard shell model, $E_{gs}=-135.94$ MeV, for the value of the cut off parameter $\eta$ we only have 
some general idea that it should be around 3 \cite{jtpden,jdenrc}.
In Fig. \ref{si28} by choosing $\eta=2.8$, the moments method reproduces quite well
the exact shell model densities.
To get a better description of the moments method level densities
we can adjust the $\eta$ parameter to optimally reproduce the exact shell-model densities.  
The cut-off parameter plays a similar role as that of the width in a Gaussian distribution. Indeed, if we increase 
the cut-off parameter then the density becomes wider and lower, while decreasing it
leads to a narrowing of the density. Fig. \ref{eta} helps to
determine the optimal value of the cut-off parameter $\eta$. In this figure the vertical axis presents
an error factor $f_{err}$, which is a measure of the deviations of the calculated density $\rho_{mm}$ 
(using the moments method) from the exact shell-model level density, $\rho_{sm}$. One possible way
to construct this error factor is the following \cite{nic10,msmh-epl}:
\be\label{ferr}
f_{err}=\exp\left(\sqrt{\frac{1}{N_i} \sum_{i=1}^{N_i} \ln^2
\left[ \frac{\rho_{mm}(E_i)}{\rho_{sm}(E_i)} \right]} \right)-1,
\ee
where the sum over $i$ spans an energy region, for which one wants to compare the level densities.
%whichare in general not considered in our model spaces.
The moments method is known to be statistically valid when the fluctuations can be described by a Gaussian
orthogonal Ensemble. Strictly imposing this condition may not be very practical. Therefore we consider
it valid in the regime where the density is at least 5-10 levels per MeV. This condition
can be used to establish a starting energy for the sum in Eq. (\ref{ferr}). 
The sum should also be upper limited to excitation energies for which the $2 \hbar \omega$ states
are not contributing significantly. For most of the model spaces considered here this upper limit
is about 10-15 MeV.
Figure \ref{eta} presents the dependence of the $f_{err}$ on the cut-off parameter $\eta$. 
It suggests that optimal values for $\eta$ are in the interval 2.5-3.0, which supports our 
initial guess. It also indicates that there is relatively small sensitivity to this parameter in
the indicated interval. Therefore, for the $pf$ and $pf+g_{9/2}$ spaces we chose $\eta=2.6$, the value for which
the moments method level densities reproduce quite well the exact shell-model level densities. 
The minimum value of $f_{err}=0.1-0.2$ offers an estimated average accuracy of the moments method for the model
spaces and the nuclei shown in the inset. A study of an optimal $\eta$ parameters for a larger class of
nuclei and model spaces will be published elsewhere \cite{msmh-epl}. One should also mention that the exact 
spin- and parity-dependent shell-model densities were calculated with the NuShellX code \cite{nushellx}.

Next we present a couple of examples for the $pf$ shell, for which we used the GXPF1A interaction \cite{gx1ap,gx1a}. 
Figures \ref{fe52} and \ref{cr52} present the results for ${}^{52}$Fe ($J=0,1$) and for ${}^{52}$Cr ($J=0,1,2,3$) 
in the $pf$ shell, for which we have used the GXPF1A interaction \cite{gx1ap,gx1a}.
The corresponding ground state energies are known and the cut-off parameter 
was chosen to be $\eta=2.6$.
One can only compare the lowest parts of the 
level densities (up to 200 levels). 
For higher excitation energies it already becomes too difficult to calculate the exact shell-model
densities, because of the large number of states needed. As one can see, the
moments method densities are in a very good agreement with the exact shell-model densities.
Figures \ref{fe52} and \ref{cr52} also includes results obtained by Goriely, et al. 
 using the Hartree-Fock-Bogoliubov (HFB) single particle energies and the 
combinatorial method \cite{gor-icomb-08,gor-comb-htm}.

%new insert
%From the data of Goriely's group one can clearly see the oscillations, which are typical for mean %field calculations. Existing of such oscillations is connected to the shell structure of the mean %field and to the fact of neglecting many-body correlations. After we turn on the residual %interaction and take into account the many-body correlations the oscillations disappear.  

%Part II. Large cases: ${}^{68}$Se and ${}^{64}$Ge.\\

\begin{figure*}[h]
\includegraphics[width=0.47\textwidth]{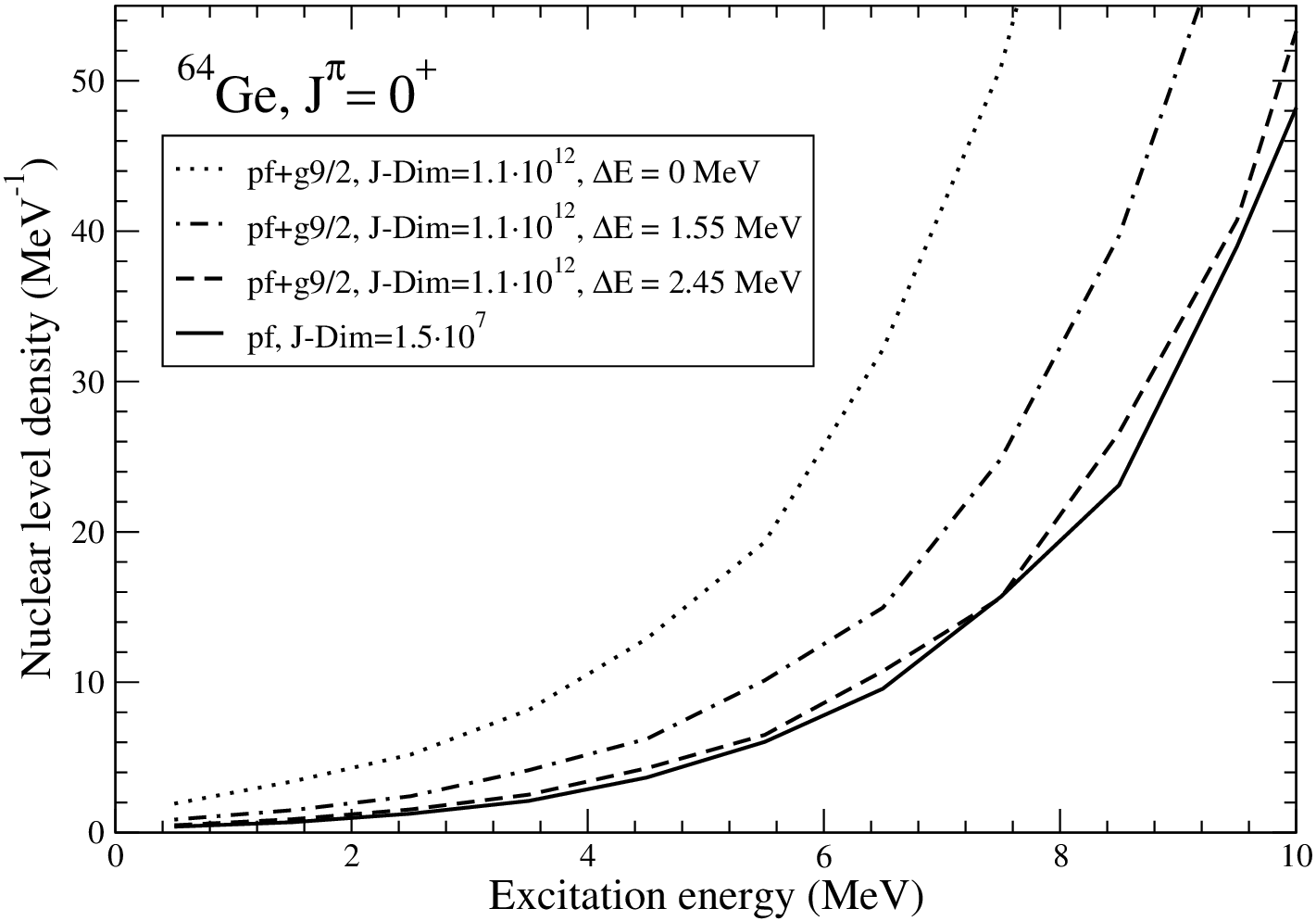}
\hfill
\includegraphics[width=0.47\textwidth]{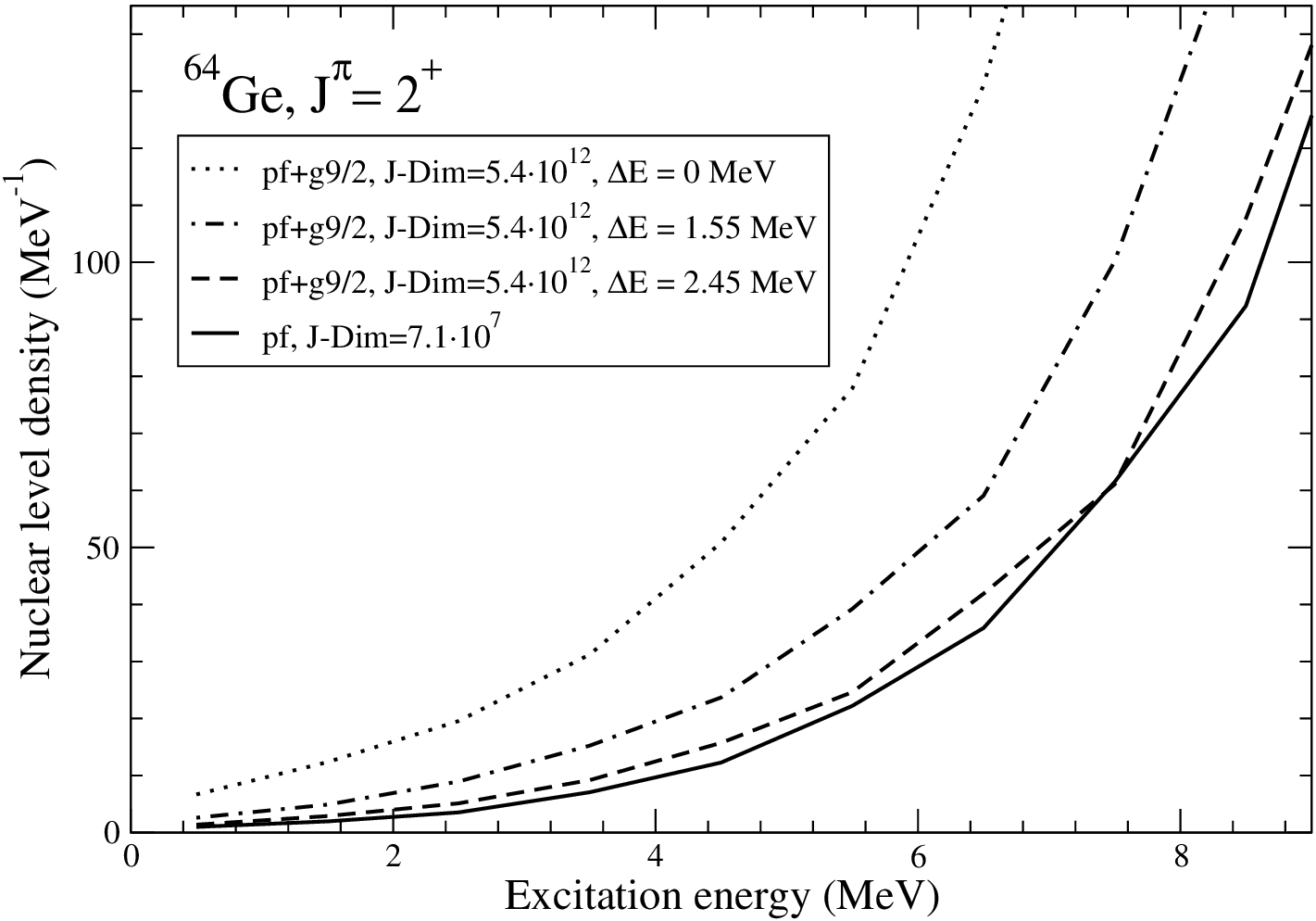}
\\
\caption{${}^{64}$Ge nuclear level densities for $J=0,2$ and positive parity. Solid line presents the
calculation in the $pf$ shell with GXPF1A interaction. For this calculation we know the ground state 
energy $E_{gs}(pf)=-304.25$ MeV. Other three lines present calculations in the large model space, when 
level $g_{9/2}$ is added. The ground state energy for these cases $E_{gs}(pf+g_{9/2})=E_{gs}(pf)-\Delta E$, 
where $\Delta E$ is the energy shift. The cut-off parameter is $\eta=2.6$.}\label{ge64}
\end{figure*}

\begin{figure*}
\includegraphics[width=0.47\textwidth]{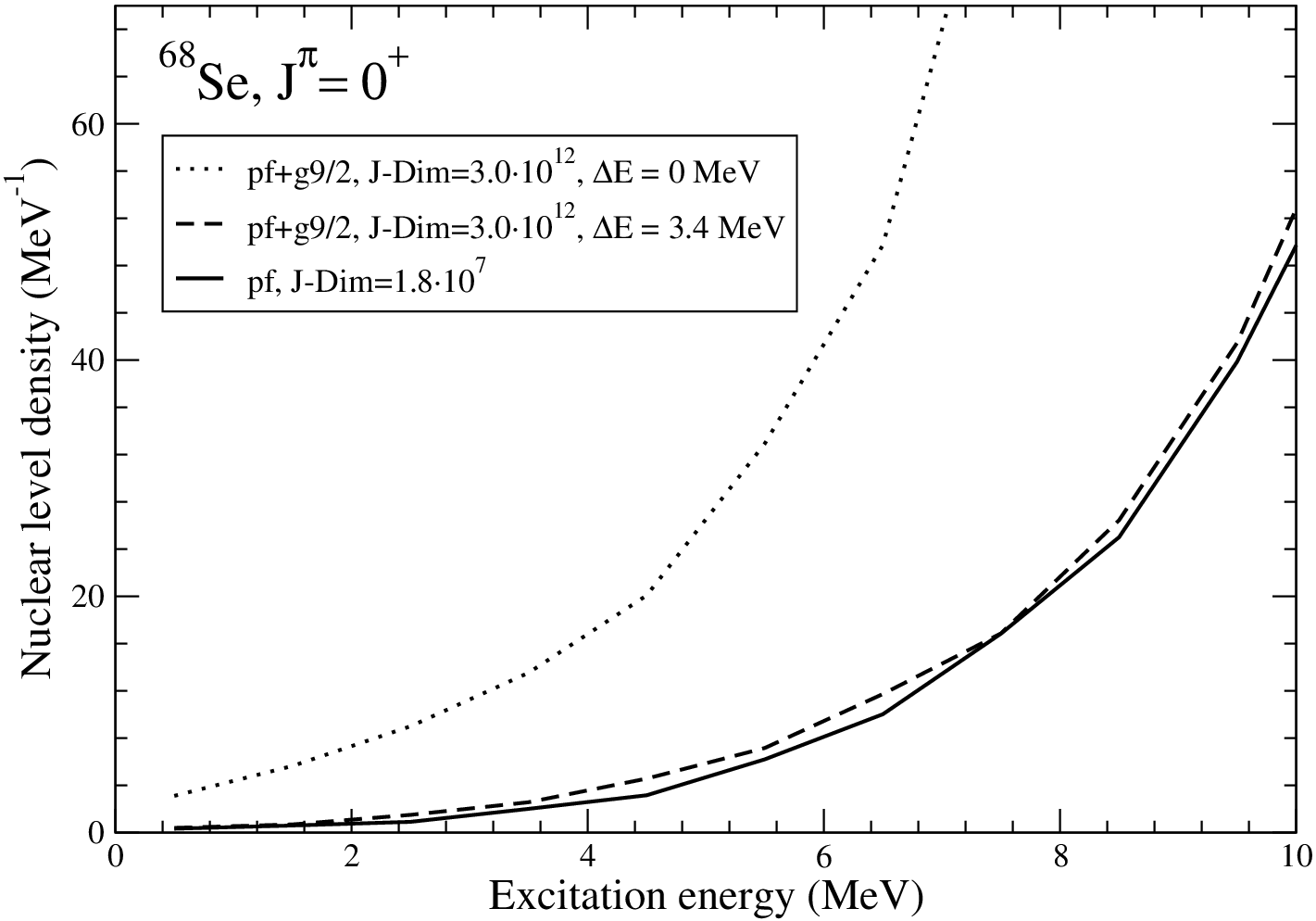}
\hfill
\includegraphics[width=0.47\textwidth]{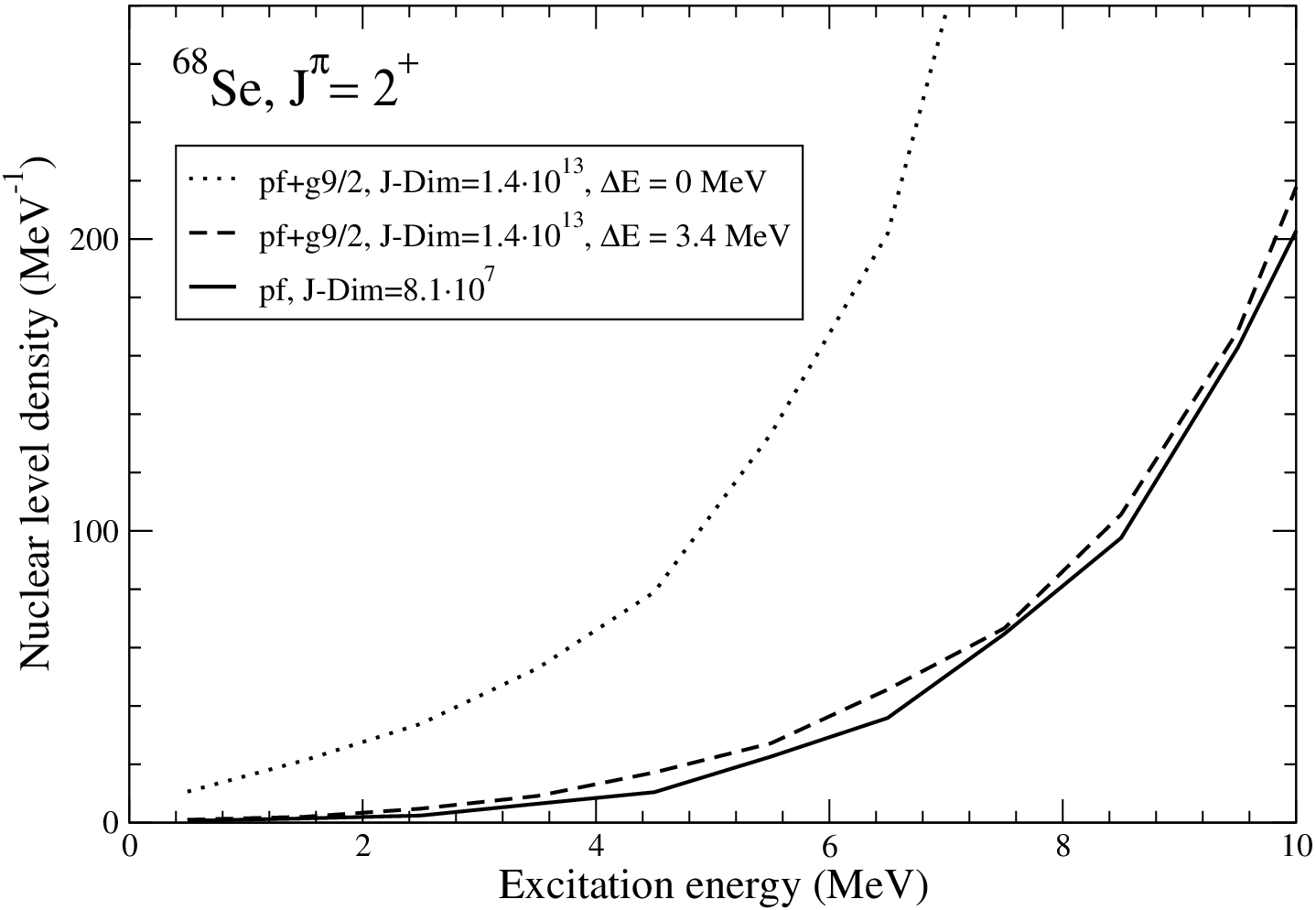}
\\
%\parbox[t]{0.47\textwidth}{ \caption{ Se68 J=0 }\label{se680}}
%\hfill
%\parbox[t]{0.47\textwidth}{\caption{ Se68 J=2 }\label{se682}}
\caption{${}^{68}$Se nuclear level densities for $J=0,2$ and positive parity. Solid line presents the
calculation in the $pf$ shell with GXPF1A interaction. For this calculation we know the ground state 
energy $E_{gs}(pf)=-353.1$ MeV. Other two lines present calculations in the large model space, when 
level $g_{9/2}$ is added. The ground state energy for these cases $E_{gs}(pf+g_{9/2})=E_{gs}(pf)-\Delta E$, 
where $\Delta E$ is the energy shift. The cut-off parameter is $\eta=2.6$.}\label{se68}
\end{figure*}

We mentioned in the introduction that one can envision using information
from level densities to extract with good approximation values for the g.s. energies. 
%Let us present now some new results we have obtained. 
Using our algorithm and
the moments method one can easily calculate the nuclear level density 
for any nucleus that can be described in the $pf+g_{9/2}$ model space. 
The interaction we used for this
model space was built starting with the GXPF1A interaction for the $pf$ model space, to which
$G$-matrix elements that describe the interaction between the $pf$ orbits and $g_{9/2}$ orbit
were added. The single particle energy for the $g_{9/2}$ orbit was fixed at -0.637 MeV. 
The calculation time for the worst case
is more than reasonable: it takes about three hours for 16 processors and only a
few minutes for 1000 processors. 
Figures \ref{ge64} and \ref{se68} summarize the results obtained for ${}^{68}$Se and ${}^{64}$Ge, nuclei that are 
believed to be "waiting-points" in the rp-process path \cite{schatz-pr98,wp07-1,wp07-2}. 
We have only presented the densities for $J=0,2$, and positive parity.

It is important to notice that in the $pf$ model space the shell-model calculations of the g.s. energies can be done. 
For the $pf$ shell we have the following ground state energies: $E_{gs}(pf)=-304.25$ MeV for ${}^{64}$Ge and
$E_{gs}(pf)=-353.1$ MeV for ${}^{68}$Se. Using these ground state energies and 
 the cut-off parameter $\eta=2.6$, we are able to calculated the nuclear level densities
according to Eqs. (\ref{rho1}) and (\ref{frg}). The solid lines on Figs. \ref{ge64} and \ref{se68}
present the densities in the $pf$ shell. 
To calculate the same level densities in the $pf+g_{9/2}$ model space we have
 to adjust the ground state energies and the cut-off parameter for this space. 
For the cut-off parameter we use the same value, $\eta=2.6$. The dotted lines show the nuclear level
densities if we keep the ground state energies as they were in the $pf$ shell.
It is natural to expect only small differences between the level densities calculated in those
two model spaces at low excitation energy. 
%In our particular case, the "low-lying" covers excitation energies up to 3-5 MeV. This region was chosen
%based on the distance of the $g_{9/2}$ level from the previous shell. 
The difference must be compensated by the fact that the ground state energy for the larger 
model space, that is $pf+g_{9/2}$, must be lower compared to the ground state energy for the smaller model space, 
that is, $pf$.
By decreasing the ground state energies for the $pf+g_{9/2}$ model space, one gets the dashed lines
on Figs. \ref{ge64} and \ref{se68}. The dash-dotted lines on Fig. \ref{ge64} correspond to an
ground state energy $E_{gs}=-305.8$ MeV of ${}^{64}$Ge, which was obtained by a truncated shell model calculation
in which up to six particles were excited from the $f_{7/2}$ orbits and/or into the $g_{9/2}$ orbit. The
$m$-scheme dimension in this calculation, $13.5\times10^{9}$, is at the upper limit of the state-of-the-art 
shell model calculation. As one can see, this value does not describe satisfactorily the low excitation
energy level densities.
 In order to make the low-lying part of the two densities very close, one has to adjust the
ground state energies for the $pf+g_{9/2}$ model space to the following values: 
\begin{eqnarray}
\label{enge}
E_{gs}(pf+g_{9/2})=-306.7 \; \mbox{ MeV for } \; {}^{64} \; \mbox{Ge}, \\
\label{ense}
E_{gs}(pf+g_{9/2})=-356.5 \; \mbox{ MeV for } \; {}^{68} \; \mbox{Se}.
\end{eqnarray}
The "low-lying part of the density" should be chosen such that the excitations 
to the $g_{9/2}$ orbit do not give a significant contribution. For these cases we use 
the interval 3-6 MeV in excitation energy.
We conclude that the g.s. energy adjustment of Eqs. (\ref{enge}) and (\ref{ense}) can be treated 
as a method of estimating the ground state energies in larger spaces. Therefore, one
can formulate now the following recipe: {\it to get the ground state energy for
a nucleus in a large model space, in which the direct shell model calculation is presently impossible, one can
calculate the nuclear level densities in the large model space and in an associated smaller model
space, for which the ground state energy calculation is possible. Then, the ground
state energy for the larger model space can be estimated by demanding that the level densities in 
the two model spaces at low excitation energy be the same or very close.} 
Certainly, one should not arbitrarily select the larger model space and the associated solvable 
model space. What we proved here seems to be valid when adding one more single particle level to a
solvable model space, such that the entire shell structure is not significantly distorted.
%Such method has some disadvantages connecting to the
%cut-off parameter uncertainty and poor theoretical basis of the moments method. 

\section{Conclusions and Outlook}

In conclusion, we developed a very efficient algorithm to calculate the shell-model spin- and parity-dependent  
configurations centroids and widths, which can be used to calculate nuclear level densities.
The new algorithm takes advantage of the separation of the model space in neutron and proton subspaces.
This separation provides two important advantages: (i) the exponentially exploding dimensions and propagators 
can be calculated more efficiently in proton and neutron subspaces, and the full results can
be recovered via simple convolutions; (ii) the number of configurations is significantly increased in the
proton-neutron formalism, which very much improves the scalability of the algorithm on 
massively parallel computers. Our tests indicate almost perfect scaling for up to 4000 cores, and
we are convinced that it can scale well up to tens of thousand of cores. 
The new algorithm is so fast that the bottleneck of the calculation is now that of the ground state
energy. That is why we cannot test our algorithm for cases that take more than 1 minute on 
4000 cores.

Therefore, we investigated the possibility of using the calculated shapes of the nuclear level
densities to extract the g.s. energy. We showed that by slightly incrementing the model space, and imposing
the condition that the level density does not change at low expectation energy, one can reliably
predict the g.s. energy, and further the full level density. This new method of extracting the
shell model ground state energy for model spaces whose dimensions are unmanageable to direct
diagonalization opens new opportunities for calculating shell model nuclear level densities of
heavier nuclei of interest for nuclear astrophysics, and nuclear energy and medical physics applications.
In particular, one can envision using effective interactions extracted from $ab-initio$ theories, such as
the G-matrix with core polarization, with some adjustable monopole corrections that can be tuned
to describe the effect of the correlations to nuclear level densities of heavy nuclei. This class of 
effective interactions is much larger than the class of pairing plus multipole interactions that 
Monte Carlo methods \cite{alhassid09} can use.

The present method could be also used in more than one major harmonic oscillator shell for medium-mass
nuclei to describe level densities of both parities. The center-of-mass spurious states could be
eliminated in these cases using a method proposed in Ref. \cite{dencom}. This method requires an 
extension of the present algorithm that will enforce restrictions on the classes of configurations
included in the widths formula, similar to the one proposed in Ref. \cite{jdenrc}. Work in this
direction is in progress.

Our method seems to exhibit some sensitivity to the cut-off parameter $\eta$. In the cases we studied
a value of about 2.8 seems to provide very good results, but further investigations
of the optimal values of this parameter are necessary. In addition, one should consider going
beyond the two-moments approach for the configuration distributions. These higher moments were used
in the past for the density of states. The $J$-dependent higher moments are more difficult to calculate,
but given the computational advances we made with the first two moments, one will envision an
efficient algorithm to calculate the higher moments in the near future.

\section{Acknowledgemnets}

The authors would like to acknowledge the DOE UNEDF grant No. DE-FC02-09ER41584 for support.
M.H. acknowledges support from the NSF grant No. PHY-0758099. 
The authors are also grateful to Vladimir Zelevinsky for useful discussions.

\end{document}